\documentclass[12pt]{iopart}
\usepackage{bm}
\usepackage{iopams}
\usepackage{graphicx}
\usepackage{color}

\begin{document}
\title{Controlling cell-matrix traction forces by extracellular geometry}

\author{Shiladitya Banerjee$^1$ and M. Cristina Marchetti$^{1,2}$}

\address{$^1$ Department of Physics, Syracuse University, Syracuse New York, 13244-1130, USA}
\address{$^2$ Syracuse Biomaterials Institute, Syracuse University, Syracuse New York, 13244-1130, USA}
\ead{mcm@phy.syr.edu}

\begin{abstract}
We present  a minimal continuum model of strongly adhering cells as active contractile isotropic media and use the model to study the effect of the geometry of the adhesion patch in controlling the spatial distribution of traction and cellular stresses. Activity is introduced as a contractile, hence negative, spatially homogeneous contribution to the pressure. The model shows that patterning of adhesion regions can be used to control traction stress distribution and yields several results consistent with experimental observations. Specifically,  the cell spread area is found to increase with substrate stiffness and an analytic expression for the dependence is obtained for circular cells. The correlation between the magnitude of traction stresses and cell boundary curvature is also demonstrated and analyzed. 
\end{abstract}

\maketitle

\section{Introduction}
Living cells actively sense and  respond to the physical geometry and stiffness of their environment, which in turn affects a variety of cellular processes, such as growth, differentiation, morphogenesis, spreading and motility~\cite{Discher2005}. Cell-matrix adhesion is mediated by integrin complexes, referred to as focal adhesions, that bind to specific ligands on the underlying matrix. Focal adhesions are mechanically linked to the actomyosin cytoskeleton inside the cell, that in turn generates contractile forces on the extracellular matrix. The interplay between substrate stiffness, intracellular contractility and the extracellular adhesion forces controls  the cell morphology and its mechanical behavior. For instance, cells adhering to soft substrate are generally found to spread less and have round morphology, while cells on stiff substrates have greater spread area with more branched shapes~\cite{Yeung2005}. 

Powerful techniques have been developed in recent years to measure the traction forces exerted by adherent cells on compliant substrates~\cite{Harris1980}.  Traction Force Microscopy is used to probe the traction stresses exerted by cells on continuous elastic gels. The stresses are inferred from measurements of the  displacements of fiducial markers embedded in the gel before and after cell detachment~\cite{Dembo1999,Butler2002}. In a  second technique  cells  plated on microfabricated pillar arrays induce bending of the elastic micropillars. The traction forces are then obtained from by assuming a linear Hooke's law relation between the measured  bending and the forces ~\cite{Tan2003}. These experiments have demonstrated that the mechanical response of adherent cells is controlled by a complex interplay of substrate stiffness and geometry, myosin activity and extracellular matrix proteins. 
Adhesive micro patterning has also been used as a tool for both controlling cell shape and study the interplay between shape and cytoskeletal organization and architecture~\cite{Thery2010}. These studies have shown that when strongly adhesive patterns force the cell boundary to exhibit regions of high curvature, traction stresses tend to be concentrated in these regions, while stress fibers develop along cell boundaries linking non-adhesive zones, confirming the crucial role of the cytoskeletal contractility and architecture in controlling cellular stresses and morphology~\cite{Rape2011}.

The role of  adhesion geometry in controlling traction force distribution  has been addressed theoretically using network models and continuum mechanical models. While  models of  continuum mechanical elements coupled to bio-chemical agents have been used before to describe the traction force distribution by adherent cells~\cite{deshpande2006}, continuum minimal models  inspired by thermoelasticity~\cite{Edwards2011} or active gel theory~\cite{Banerjee2011} have recently provided new key analytical results. Network models of the contractile cytoskeleton have also been used to describe the relation between force distribution and shape of adherent cells~\cite{lemmon2010,Torres2012}, including networks of Hookean springs as well as cable networks that incorporate the asymmetry of the elastic response of biopolymers such as filamentary actin to compression version extension, with and without the explicit inclusion of contractility. In particular, the active cable network reproduces the arc morphology of cell boundaries pinned by strong local adhesions that has been seen in experiments~\cite{Lemmon2005}. The relationship between cell shape and adhesion geometry has also been studied by modeling cells as contractile films bounded by the elastic cortex~\cite{Barziv1999,Bischofs2009,Banerjee2012b}. In this paper we consider a continuum model of cells as linear, active elastic media and demonstrate that the introduction of  activity  as a \emph{spatially homogeneous} contractile, hence negative, contribution to the pressure is sufficient to reproduce the spatial inhomogeneous distribution of traction and cellular stresses observed in experiments for a number of cell geometries. An interesting extension of our work will be to introduce nonlinearity in the continuum model to incorporate an asymmetric response to compression and stretching. This asymmetry, arising from the nonlinear force-extension curve of actin filaments, is known to be important in controlling the contractile behavior of isotropic gels~\cite{Liverpool2009,Banerjee2011b} and may alter the stress distribution in adhering cells. 

In the next section we introduce our continuum  model of adherent cells as active contractile elastic media. We then use the model to study the effect of the geometry of the adhesion region on controlling the spatial distribution of stresses in the cell. The model can be solved analytically for a circular cell, where we obtain an expression for the cell spread area as a function of substrate stiffness and show that our results compares favorably to experiments (inset of Fig.~\ref{fig:spreading}).
The cases of elliptical, square and triangular cells are solved numerically. We show that the geometry of the adhesive region strongly affect the stress distribution, with traction stresses concentrated in regions of high curvatures or at sharp corners (Fig.~\ref{fig:2dshapes}). In section 3.3 we provide an analytical argument that quantifies the correlation between traction stress magnitude and curvature of the cell boundary and discuss in section 3.4 the relative roles of shear and compressional deformations in controlling the stress distribution. We conclude with a brief discussion.

\section{Adherent cell as a contractile gel}
We consider a stationary cell adhering to an elastic substrate via stable focal adhesion complexes. We further assume that the cell has attained its optimum spread area on the substrate, with an average height $h$ much thinner than its perimeter. In mechanical equilibrium, the condition of local force-balance translates to $\partial_\beta \sigma_{\alpha \beta}=0$, where ${\bm \sigma}$ is the three-dimensional stress tensor of the cell with greek indices taking values $x,y$ and $z$. For a thin cellular film we average the cellular force-balance equation over the cell thickness $h$. In-plane force balance is given by
\begin{equation}
\label{eq:balance-2d}
\partial_j\sigma_{ij} + \partial_z \sigma_{iz}=0\;,
\end{equation}
with $i, j$ denoting in-plane coordinates. We assume that the top surface of the cell is stress free, $\sigma_{iz}({\bf r}_\perp,z=h)=0$, whereas at the cell-substrate interface $z=0$, the cell experiences lateral traction stresses given by $\sigma_{iz}({\bf r}_\perp,z=0)=Yu_i({\bf r}_\perp,z=0)$. Here, $Y$ denotes the substrate rigidity parameter, representing the cell-substrate anchoring strength, and ${\bf u}({\bf r}_\perp,z)$ is the in-plane deformation field of the cellular medium.   The thickness-averaged force balance equation then reads~\cite{Banerjee2011,Edwards2011},
\begin{equation}\label{eq:force-balance}
h\partial_j\overline{\sigma}_{ij}=Yu_i\;,
\end{equation}
where $\overline{\sigma}_{ij}({\bf r}_\perp)=\int_0^h(dz/h)\sigma_{ij}({\bf r}_\perp,z)$. It is worthwhile to mention that the assumption of in-plane traction forces is a good approximation for fully spread stationary cells making almost zero contact angle with the substrate. During the early stages of spreading and migration, cells can exert appreciable out-of-plane traction forces via rotation of focal adhesions~\cite{Legant2013}. In the following we will drop the overbear indicating the average and refer to thickness averaged quantities throughout.
The quantity $T_i=Yu_i$ is a stress in three dimensions, i.e., a force per unit area. It describes the in-plane traction force per unit area that the cell exerts on the substrate. 
The assumption of local elastic interactions with the substrate strictly holds on elastic substrates that are much thinner than the lateral size of the cell~\cite{Banerjee2012} or on micropillar substrates~\cite{Edwards2011}. The substrate rigidity parameter $Y$ depends on the stiffness of the underlying substrate as well as on the density $\rho_f$ and stiffness $k_f$ of focal adhesions. For an elastic substrate of shear modulus $\mu_s$ and thickness $h_s$, $Y$ takes the simple form~\cite{Banerjee2012}, $Y^{-1}=\frac{1}{k_f \rho_f} + \frac{1}{\mu_s/h_s}$. 

We model the cell as an isotropic and homogeneous elastic material with additional internal active stresses due to actomyosin contractility. The constitutive relation for the cellular stress tensor is then given by,
\begin{equation}\label{eq:stress}
\sigma_{ij}=\frac{E}{2(1+\nu)}\left(\frac{2\nu}{1-2\nu}\bm\nabla\cdot{\bf u}\ \delta_{ij} + \partial_j u_i + \partial_i u_j \right) + \sigma_a \delta_{ij} \;,
\end{equation}
where $E$ and $\nu$ denote the Young modulus and Poisson ratio  of the cellular material, respectively. Actomyosin contractility is modeled as a negative contribution to the local pressure, corresponding to  $\sigma_a>0$. The assumption of linear elasticity is valid on time scales shorter than cytoskeletal turnovers, that are indeed slowed down by strong adhesion to the substrate. Equations \eref{eq:force-balance} and \eref{eq:stress}, subject to the boundary condition $\sigma_{ij}n_j \vert_\Omega=0$, wtih $\Omega$ the cell boundary and ${\bf n}$ the outward unit normal on $\Omega$, completely describe the equilibrium of an adherent cell. As a consequence of the stress free condition at the lateral cell boundary, the net traction force transmitted by the cell to the substrate vanishes, i.e., $\int_A d^2{\bf r}\ Yu_i=\oint_\Omega ds\  \sigma_{ij}n_j=0$. It is instructive to consider two limiting cases for the anchoring strength. When the cell is rigidly anchored onto the substrate, corresponding to $Y\rightarrow \infty$, we find ${\bf u}=0$, defining the reference state for elastic deformations. In our model the reference cell shape is then dictated by the geometry of the adhesion patch, which can be controlled in experiments by micropatterning substrates by adhesion proteins. In contrast, when $Y\rightarrow 0$, the cell does not adhere to the substrate  and the equilibrium state is uniformly contracted state, with a density enhancement $\delta\rho=-\bm\nabla\cdot{\bf u}=\sigma_a(1+\nu)(1-2\nu)/E(1-\nu)$. In the following, we investigate analytically and numerically solutions of the cell elasticity equations \eref{eq:force-balance} and \eref{eq:stress} subject to stress-free boundary conditions in various planar geometries.

\section{Results}
\subsection{Spatial distribution of traction stresses is sensitive to adhesion geometry}
\begin{figure}
\centering
\includegraphics[width=0.6\textwidth]{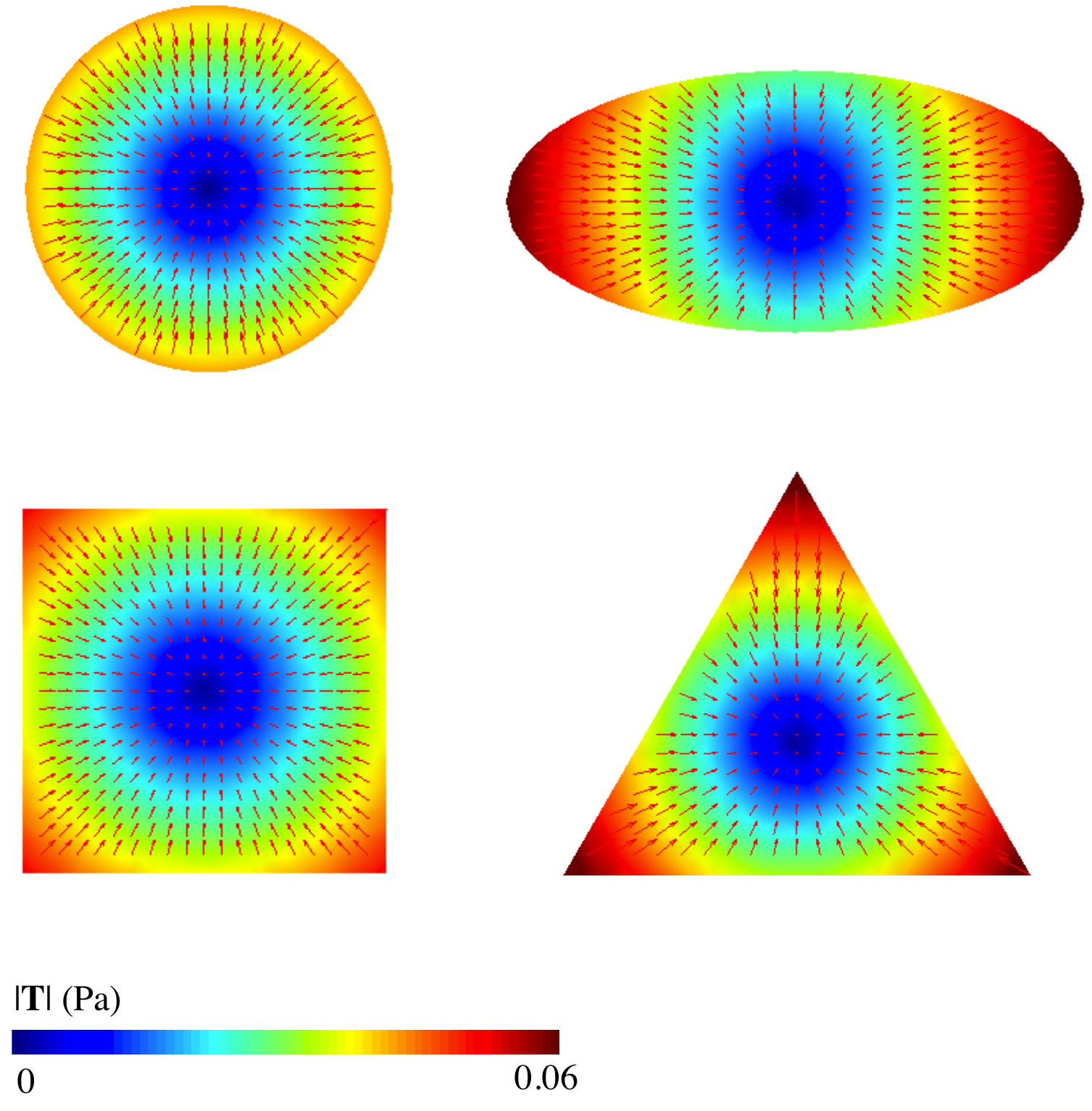}
\caption{\label{fig:2dshapes}Equilibrium cell shapes for various adhesion patterns : Circle (top left), ellipse (top right), square (bottom left) and equilateral triangle (bottom right).  The color map indicates magnitude of the traction  $\vert {\bf T}\vert=Y\vert{\bf u}\vert$, and the arrows demote the direction of the traction vectors. The reference shapes for all the four patterns have an equal area of $1000\ \mu m^2$. The other parameters are: $E=1~{\rm kPa}$, $\nu=0.4$, $\sigma_a=1~{\rm kPa}$, $\mu_s=10~{\rm kPa}$, $h_s=30~ \mu{\rm m}m$, $h=0.2~ \mu{\rm m}$.}
\end{figure}
The spatial distribution of traction stresses exerted by cells on substrate and the corresponding  organization of stress and deformation inside the cell are affected by the geometry of adhesive patterns. Using micropatterning techniques, cell shapes can be constrained to adhere to controlled geometrical patterns~\cite{Chen1997,Thery2006}. In our model the shape determined by the pattern in the limit of infinite adhesion strength provides the reference shape for the cell. Here we  investigate four  reference cell shapes: circle, ellipse, square and equilateral triangle. These are chosen to have the same reference area but different perimeters.  The case of a circular cell can be treated analytically, as described below. For the other shapes the elasticity  equations~\eref{eq:force-balance} and \eref{eq:stress} are solved numerically using the  \textsc{MATLAB} pde toolbox. We assume the  contractility $\sigma_a$ to be uniform and of order of the cellular Young's modulus. Heatmap of traction stresses are shown in Fig.~\ref{fig:2dshapes}. In all cases the traction stresses are concentrated at the cell periphery, irrespective of the reference shape. The magnitude of the local traction stress is, however,  higher in regions of high curvatures or at sharp corners.

For a circular cell, Eqs.~\eref{eq:force-balance} and \eref{eq:stress} can be solved  analytically~\cite{Edwards2011,Mertz2012}. Assuming in-plane rotational symmetry, it is convenient to use polar coordinates $r$ and $\theta$, denoting radial and angular coordinates, and demand that no quantity depend on $\theta$. The equation for the radial displacement $u_r$ about a circular reference state of radius $R_0$, is then given by
\begin{equation}
\label{eq:ur}
r^2\partial_r^2 u_r + r\partial_r u_r - (1+r^2/\ell_p^2) u_r =0\;,
\end{equation}
where the \textit{penetration length} $\ell_p$ describes the localization of traction stresses at the cell boundary. It is given by :
\begin{equation}
\label{eq:lp}
\ell_p^2=\frac{Eh(1-\nu)}{Y(1+\nu)(1-2\nu)}\;,
\end{equation}
and is essentially controlled by the ratio of the cell stiffness $\sim E$ to the substrate rigidity $\sim Y$. The penetration length is short on stiff substrates and increases with decreasing substrate rigidity.
The solution of Eq.~\eref{eq:ur} with the boundary conditions $\sigma_{rr}(r=R_0)=0$ and $u_r(r=0)=0$ is given in terms of modified Bessel functions of the first kind as,
\begin{equation}\label{eq:circle}
u_r(r)=-\sigma_a R_0\left[\frac{(1+\nu)(1-2\nu)}{E(1-\nu)}\right]I_1(r/\ell_p)g(R_0/\ell_p)\;,
\end{equation}
with $g(s)=\left[s I_0(s)-\frac{1-2\nu}{1-\nu}I_1(s)\right]^{-1}$. As anticipated, the deformation $u_r$ vanishes for all $r$ when $Y\rightarrow\infty$, when the adhering circular cell is maximally spread and has its largest undeformed radius $R_0$.

\subsection{Cell spread area is sensitive to substrate stiffness and contractility}
The optimal spread area of the cell is controlled by the interplay between cell contractility, as described by the active pressure $\sigma_a$, and the traction forces on the substrate. In the case of a circular cell, where the deformation induced by adhesion is given by Eq.~\eref{eq:circle}, the steady state cell area is given by,
\begin{equation}
\label{eq:area}
A=\pi (R_0+u(R_0))^2\;,
\end{equation}
with $R_0$  the reference radius corresponding to the maximal spread area $A_\infty=\pi R_0^2$ attained on an infinitely rigid substrate, where $u_r(r)=0$. To make contact with experiments, we investigate the ratio $A/A_\infty$, the relative cell spread area, as a function of substrate stiffness and contractility.
\begin{figure}
\centering
\includegraphics[width=0.9\textwidth]{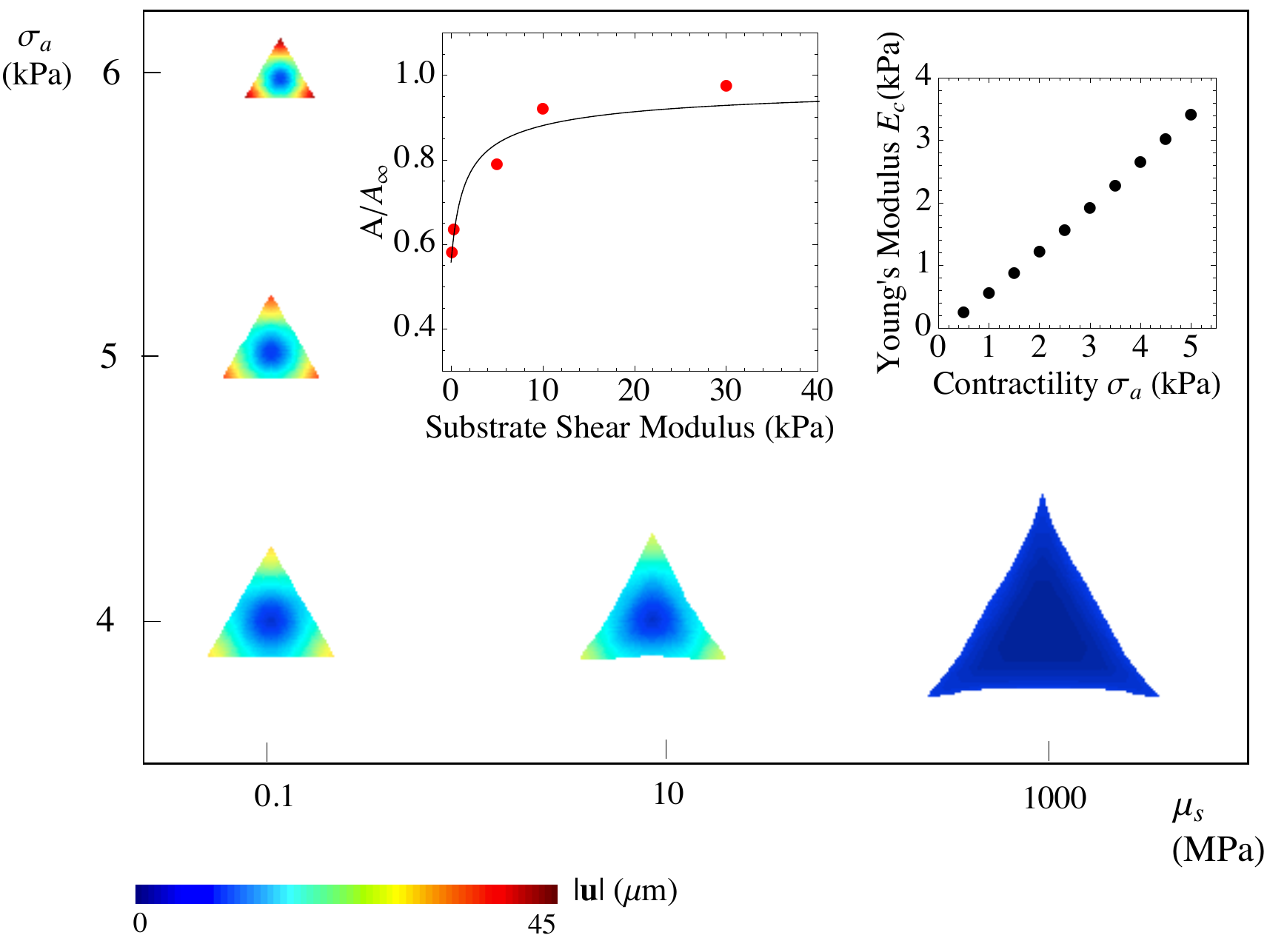}
\caption{\label{fig:spreading}Optimal shape of a triangular cell  for different values of the active pressure $\sigma_a$ and the substrate shear modulus $\mu_s$, with $E= 1~{\rm Pa}$. The color map represents the magnitude of the displacement vector $\vert{\bf u}\vert$ (proportional to the traction force) about an equilateral triangular reference shape of area $1000~\mu{\rm m}^2$. The cell spread area increases with increasing substrate stiffness and decreases with increasing $\sigma_a$. Inset (Left) : Least-square fit of the relative cell spread area $A/A_\infty$ obtained from the model using Eq.~\eref{eq:area} (solid) to the experimental data reported in Ref.~\cite{Chopra2011} (solid red circles). The  fitting parameters are $E=911~{\rm Pa}$ and $\sigma_a=1589~{\rm Pa}$. Inset (Right) : Relationship between cellular Young's modulus $E_c$ and contractility $\sigma_a$. Here we tune $\sigma_a$ to desired values and then determine the fitting parameter $E_c$ using data in Ref.~\cite{Chopra2011}. Other parameters : $\nu=0.4$, $h_s=30~\mu{\rm m}$, $h=0.2~\mu{\rm m}$.}
\end{figure}
On stiff substrates, where $R_0\gg \ell_p$, i.e., the traction stress extends over a length much smaller than the reference cell radius, $u_r(R_0)\simeq -\sigma_a \ell_p/B$, where the compressional modulus $B$ is given by $B=E(1-\nu)/\left[(1+\nu)(1-2\nu)\right]$. The relative spread area then takes the simple form $A/A_\infty \simeq \left(1-\frac{\sigma_a}{R_0}\sqrt{h/BY}\right)^2$. Letting $Y\simeq \mu_s/h_s$, we note that increasing substrate stiffness increases relative spread area, with $A/A_\infty \rightarrow 1$ as $\mu_s \rightarrow \infty$, in qualitative agreement with experiments~\cite{Yeung2005,Ghibaudo2008,Chopra2011}. In contrast, increasing the contractile pressure $\sigma_a$ reduces the optimal cell spread area, consistent with the experimental observation that myosin-II activity retards cell spreading~\cite{Wakatsuki2003}. To make a quantitative comparison with experiments, we fit Eq.~\eref{eq:area} to experimentally reported data on the projected area of cardiac myocytes cultured on N-cadherin coated PA gels of varying stiffnesses~\cite{Chopra2011}.  Here the maximal spread area $A_\infty$ is taken to be equal to the cell projected area on a glass substrate (shear modulus $\sim$ 30 GPa), which is $\simeq 690\ \mu m^2$. The fit, shown in the left inset of  Fig.~\ref{fig:spreading}, is obtained using the active contractility $\sigma_a$ and the cellular Young's modulus $E$ as the fitting parameters. A least-square fit gives us $E=911~{\rm Pa}$ and $\sigma_a=1589~{\rm Pa}$. Although the strength of contractility is likely to depend on cell type, it is worth highlighting that the fit value for $\sigma_a$ is of the same order of magnitude as previously used in Ref.~\cite{Mertz2012} to fit the measured value of the surface tension of a colony of epithelial cells. Next, we tune the contractility $\sigma_a$, which can be artificially controlled through pharmacological interventions, and determine the corresponding best fit value of the cellular Young's modulus $E_c$. Our result (Fig.~\ref{fig:spreading}, right inset) indicates a linear relationship between the cellular Young's modulus and the contractile stress. There are indeed experimental data available~\cite{Wang2002} that show that the cell stiffness increases linearly with contractility for adherent cells~\footnote{We thank the anonymous referee for suggesting this fit and pointing out  to us Ref.~\cite{Wang2002}.}. This suggests that our model could be used to infer contractility from measurements of cellular stiffness. Figure~\ref{fig:spreading} also demonstrates the competing roles of contractility and adhesion in controlling optimal cell shapes for a chosen triangular reference state. On softer substrates the triangular cell retains its topology and contracts by an amount proportional to $\sigma_a$, whereas on stiffer substrates the corners tend to form protrusions.

\subsection{Traction forces increase with cell boundary curvature}
\begin{figure}
\centering
\includegraphics[width=0.8\textwidth]{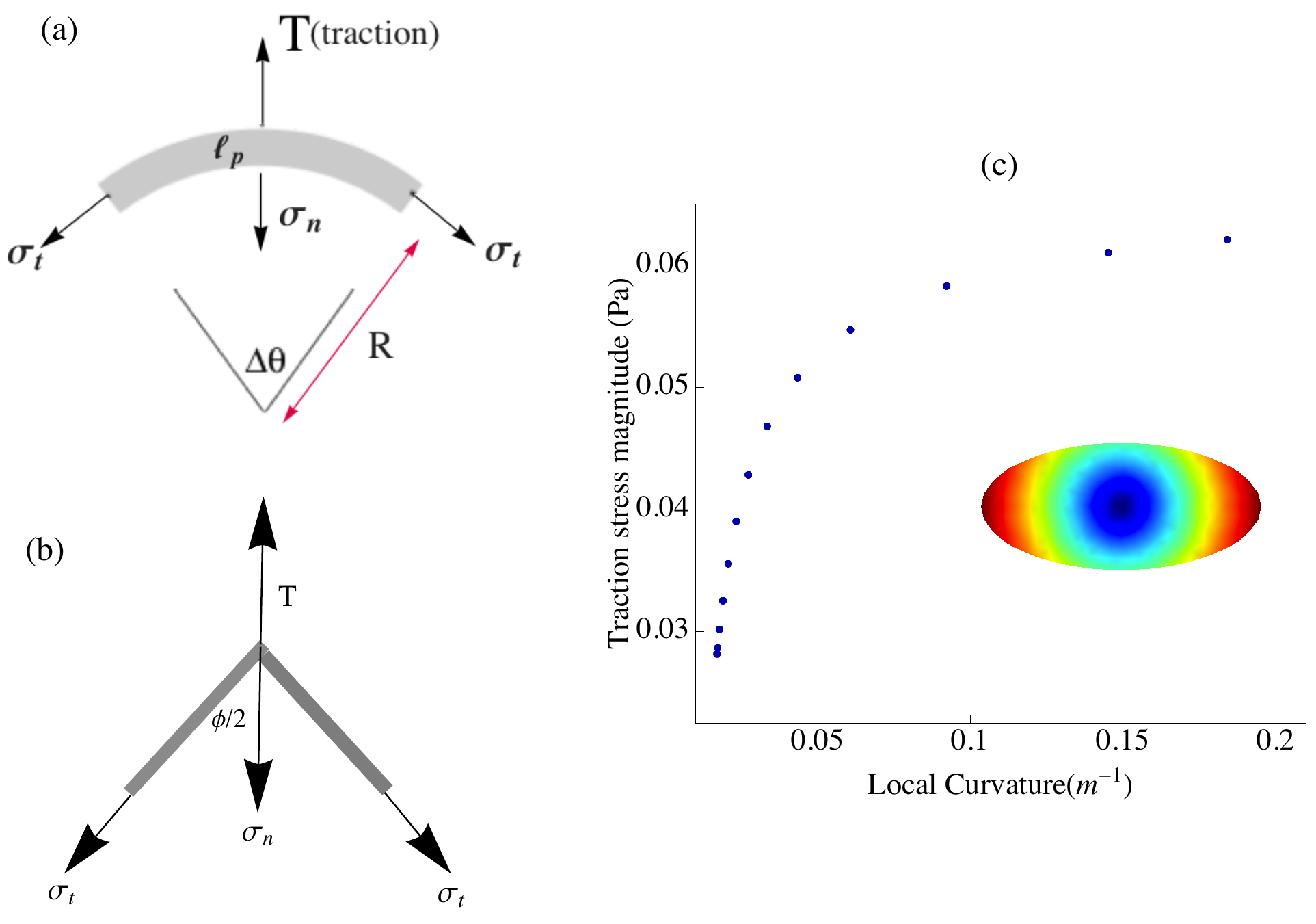}
\caption{\label{fig:curv}(a) Force-balance on a thin slice of cellular material at the cell boundary. (b) Force-balance at a generic sharp corner with opening angle $\phi$. (c) Traction stress magnitude at the cell edge as a function of the local curvature $\kappa$ for the elliptical cell of Fig.~\ref{fig:2dshapes}. }
\end{figure}
When the boundaries of the adhesion pattern exhibits non-uniform curvature, the traction stresses are higher at regions of high  curvatures. This is seen for example in Fig.~\ref{fig:2dshapes} for the case of an elliptical reference shape. To justify this we propose a simple analytical argument based on local force balance. Consider a thin slice of cellular material at the cell periphery of width comparable to penetration length $\ell_p$ and arc length $R\Delta\theta $ much less than the cell perimeter (Fig.~\ref{fig:curv}(a)), with $1/R$ the local curvature of the cell element. At the outer edge of this element, the only force on the cell is the reaction to the traction by the cell on the substrate  traction, of areal density $-{\bf T}$, with ${\bf T}=Y{\bf u}$.  This yields an outward total force on the outer edge of the cell element of magnitude $TR\Delta\theta\ell_p$, with $T>0$. At the interior edge, the cellular element experiences a contractile force of magnitude $\sigma_n (R-\ell_p)\Delta\theta\ell_p$, where $\sigma_n$ is the normal stress pulling the inner contour inwards and has contributions from active as well as passive elastic stresses. The lateral stresses $\sigma_t$ contributes to an effective line tension $\sigma _t \ell_p R\Delta\theta$ of the cell element. Due to the curvature of the boundary element, the line tension generates an inward Laplace pressure of magnitude $\sigma_t \ell_p/R$. Local balance of forces then yields,
\begin{equation}
T R\Delta\theta\ell_p-\sigma_n (R-\ell_p)\Delta\theta\ell_p=R\Delta\theta\ell_p\sigma_t\frac{\ell_p}{R}\;.
\end{equation}
The above law can be written down in a compact form as,
\begin{equation}\label{eq:curv}
T=\sigma_n + (\sigma_t-\sigma_n)\ell_p \kappa\;,
\end{equation}
with $\kappa=1/R$, the local curvature of the boundary element. Equation~\eref{eq:curv} then tells us that local magnitude of traction  increases linearly with increasing boundary curvature. The lateral and normal stresses $\sigma_t$ and $\sigma_n$ can be expressed in terms of the local cellular stresses in polar coordinates as $\sigma_t = \sigma_{\theta\theta}-\partial_\theta\sigma_{r\theta}$ and $\sigma_n=\sigma_{rr}$. The linear dependence of $T$ on $\kappa$ strictly holds in the limit $\ell_p \kappa \ll 1$. In addition, non-local elastic interactions can also affect the dependence of traction  magnitude on local curvature. Figure.~\ref{fig:curv}(c) shows the dependence of the magnitude of the traction stress at the cell boundary on local curvature for an elliptical cell as shown in Fig.~\ref{fig:2dshapes}. For low $\kappa$, the traction stress magnitude increases linearly with $\kappa$ before reaching a plateau at higher values of $\kappa$.

When the cell boundary exhibits a sharp corner with opening angle $\phi$, as shown in  Fig.~\ref{fig:curv}(b), the local force-balance is  given by,
\begin{equation}
T=\sigma_n + 2\sigma_t \cos{(\phi/2)}\;,
\end{equation}
where $\sigma_n$ acts along the bisecting line of the corner. Hence smaller the opening angle, the larger is the traction force.

\subsection{Mechanical anisotropy induced by geometric anisotropy}
\begin{figure}
\centering
\includegraphics[width=1.0\textwidth]{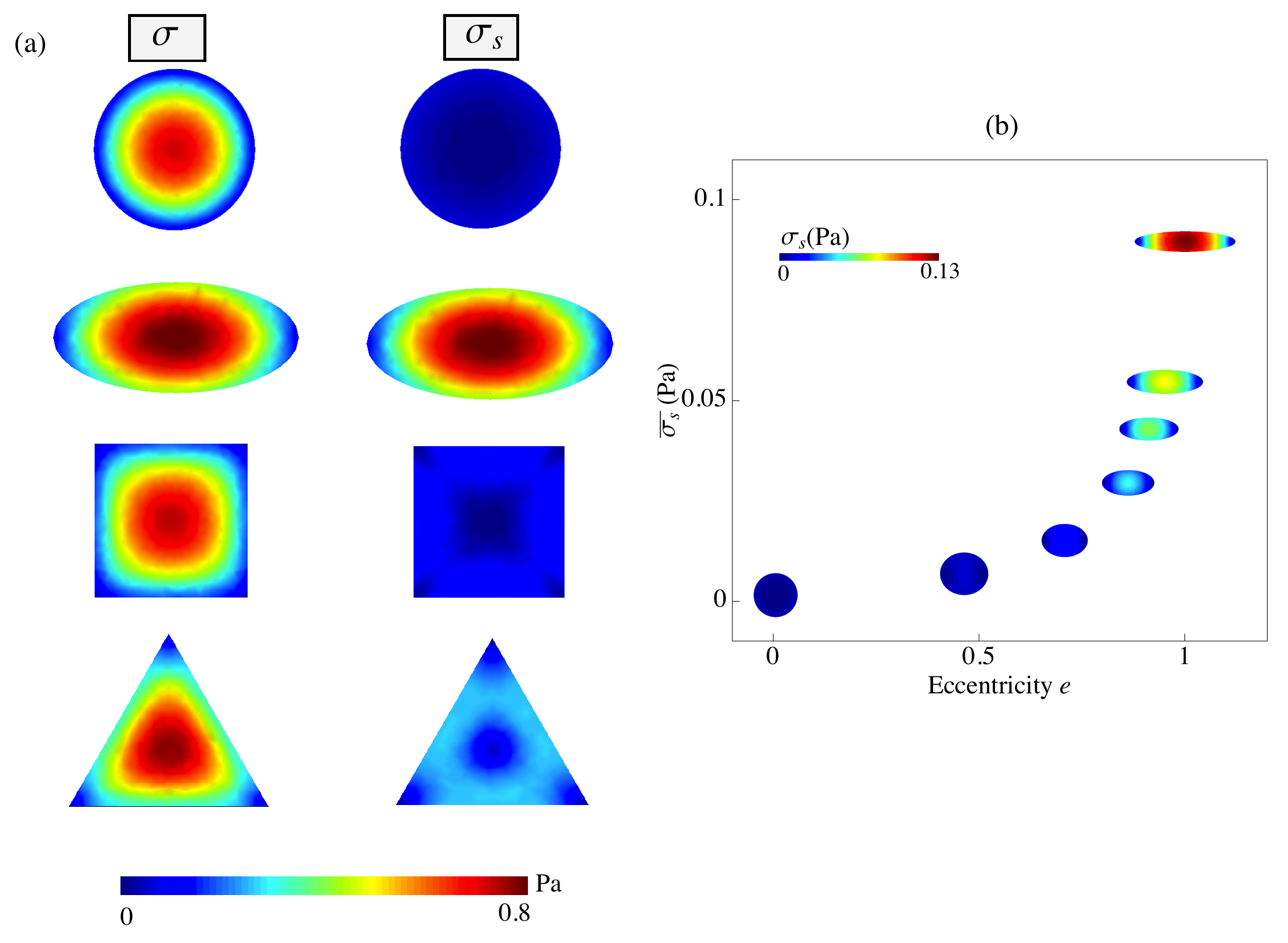}
\caption{\label{fig:stress}Cell shape anisotropy correlates with internal stress anisotropy. (a) Heatmap of internal compressive stress $\sigma$ (left) and maximum shear stress $\sigma_s$ (right) corresponding to various reference shapes : circle, ellipse, square and equilateral triangle. The reference shapes all have an equal area of $1000\ \mu m^2$. (b) Average maximum shear $\bar{\sigma}_{s}$ as a function of eccentricity $e$ for elliptical cells of same reference area ($1000\ \mu m^2$). Equilibrium shapes with colorplot of $\mu$ are given as plot markers. Parameters : $E=1$ kPa, $\nu=0.4$, $\sigma_a=1$ kPa, $\mu_s=10$ kPa, $h_s=30\ \mu m$, $h=0.2\ \mu m$.}
\end{figure}
The spatial distribution of internal stresses $\sigma_{ij}$ within the cell depends on cell shape, which is in turn controlled by the geometry of the adhesive region. Experimentally ${\bm \sigma}(x,y)$  can be obtained  from the measured distribution of traction stresses ${\bf T}(x,y)$, inverting the local force-balance condition $\partial_j \sigma_{ij}= T_i$~\cite{Tambe2011}.  The elasticity equations  Eqs.~\eref{eq:force-balance} and \eref{eq:stress} can be recast as a single partial differential equation for the internal stress tensor $\sigma_{ij}$, given by
\begin{equation}
\label{eq:internal-stress}
\ell_p^2\left[\partial_i\partial_k\sigma_{kj}\right]^S + \delta_{ij}\sigma_a = \sigma_{ij} + \frac{1-2\nu}{\nu}\delta_{ij}\left(\sigma_{kk}-2\sigma_a\right)\;,
\end{equation}
where $[...]^S$ denotes symmetrization with respect to indices that are not summed over, i.e.,
$\left[\partial_i\partial_k\overline{\sigma}_{kj}\right]^S=\frac12\left[\partial_i\partial_k\overline{\sigma}_{kj}+\partial_j\partial_k\overline{\sigma}_{ki}\right]$. We have investigated numerically the solution of Eq.~\eref{eq:internal-stress} with stress free boundary condition $\sigma_{ij}n_j=0$. To understand the role of shear and compressional deformations in different geometries, it is instructive to diagonalize the stress tensor and display the results in terms of linear combinations of the eigenvalues $\sigma_1$ and $\sigma_2$. The sum $\sigma=\frac{1}{2}(\sigma_1 + \sigma_2)$ is simply half the trace of the stress tensor and describes compressional deformations. The difference $\sigma_s=\frac{1}{2}\vert \sigma_1 -\sigma_2 \vert=\sqrt{[\sigma_{xx}-\sigma_{yy}]^2+4\sigma_{xy}^2}$, is controlled by normal stress $\sigma_{xx}-\sigma_{yy}$ and shear stress $\sigma_{xy}$. For an isotropic reference shape, such as the circle, $\sigma_1=\sigma_2$ and  $\sigma_s= 0$, whereas for anisotropic shapes such as the ellipse, one expects nonzero values for the local maximum shear $\sigma_s$.

Fig.~\ref{fig:stress}(a) shows heatmaps of the spatial distribution of $\sigma$ and $\sigma_s$ for various reference shapes - circle, ellipse, square and equilateral triangle. Irrespective of the shape of the adhesion geometry, $\sigma$ is maximum at the cell center, indicating build-up of compressive stresses. The compressional stress $\sigma$ always vanishes at the boundary, and it does so more rapidly  at regions of high curvature or at sharp corners. In contrast, the shear stress $\sigma_s$ is identically zero for isotropic shapes, defined as those that have a gyration tensor that is diagonal, with equal eigenvalues. The circle, triangle and square are all in this class. Local stress anisotropy as measured by $\sigma_s$ is nonzero for elliptical shapes and shear stresses build up at the center of the ellipse. The shape anisotropy of ellipses can be quantified by their eccentricity $e=\sqrt{1 - (b/a)^2}$, with $a$ and $b$  the semi-major and semi-minor axes. Figure~\ref{fig:stress}(b) shows the spatial average of $\sigma_s$ over the area $A$ of the cell, defined as $\bar{\sigma}_s=\frac{1}{A}\int_A d^2{\bf r}\ \sigma_s$,  as a function of the eccentricity $e$. The average shear stress  $\bar{\sigma}_s$ increases with $e$ with a sharp rise as $e\rightarrow 1$, indicating a positive relationship between geometrical and mechanical anisotropy in adherent cells. Our theoretical model thus confirms the experimental result that cell mechanical anisotropy increases with increasing aspect ratio, as previously reported for single endothelial cells with the same spread area~\cite{roca2008}.

\section{Discussion}
We have used a continuum model of an adherent cell on a substrate as an active contractile medium to study the role of adhesion geometry in controlling cell shape, cell spreading and the spatial distribution of traction stresses. More realistic future modeling should take into account that a cell is a highly heterogeneous material with spatially varying stiffness~\cite{Heidemann2004}. It is however intriguing to note that the simplified assumption of homogeneity and isotropy in the underlying cytoskeletal network can reproduce several of the known experimental results. The central input of the model is the cell contractility or activity $\sigma_a$, a negative contribution to the pressure that enters the constitutive equation for the cellular material. In general, $\sigma_a$ will be determined by the concentration and activity of myosin proteins cross linking the actin cortex and controlling the formation of stress fiber. In our model $\sigma_a$ is assumed to be a constant parameter, to be determined by fitting experiments. We consider cells adhering to flat substrates that have been patterned with adhesive patches, consisting for instance of fibronectin coatings, of specific geometry and examine the role of the geometry of the adhesive patch in controlling the spatial distribution of stresses in the cellular material. The reference state for our cell is the limit of infinitely strong adhesion, where the cell shape and lateral extent and determined entirely by the shape and size of the adhesive patch. For finite adhesion strength, cell elasticity and contractility yield deviations form this reference state. We restrict ourselves to considering continuous or densely spaced adhesion sites. For discrete or sparsely distributed adhesion sites, non-adherent segments in the cell boundary could likely exhibit morphological transitions induced by contractile activity and substrate stiffness~\cite{Banerjee2012b}.
In agreement with experimental observations, we find that cells spread more on stiff substrates and we provide an expression for the cell area versus substrate stiffness for the case of a circular cell. We show that this expression fit the data for spread areas of  cardiac myocytes on substrates of various sitffness values(see inset of Fig. 2). We demonstrate analytically and numerically that strong traction stresses correlate with regions of high cell boundary curvature, in agreement with experimental observations. Further, as reported in experiments on single endothelial cells, our model demonstrates that cell mechanical anisotropy is higher on elongated cells than on rounded ones for fixed area~\cite{roca2008}. 

Understanding the relation between cell morphology, the cell's mechanical response and cell fate is an important  question in cellular biophysics. Our simple model highlights the correlation between the geometry of adhesion sites and cell morphology and demonstrates that traction forces by cells can be tuned by controlling the geometry of adhesive regions. An important open question not addressed by this simple model where the adhesive patch geometry solely controls the cell shape is how cell morphology is determined by the interplay of cell-substrate adhesion and dynamical reorganization of the cytoskeletal architecture in response to the adhesion stimulus. To understand this it will be necessary to incorporate the dynamical feedback between actin reorganization and adhesion kinetics.

\vspace{0.2in}
This work was supported by the National Science Foundation through award DMR-1004789.

\section*{References}
\bibliographystyle{unsrt}
\bibliography{references}

\begin{thebibliography}{10}

\bibitem{Discher2005}
D.E. Discher, P.~Janmey, and Y.~Wang.
\newblock Tissue cells feel and respond to the stiffness of their substrate.
\newblock {\em Science}, 310(5751):1139, 2005.

\bibitem{Yeung2005}
T.~Yeung, P.C. Georges, L.A. Flanagan, B.~Marg, M.~Ortiz, M.~Funaki, N.~Zahir,
  W.~Ming, V.~Weaver, and P.A. Janmey.
\newblock Effects of substrate stiffness on cell morphology, cytoskeletal
  structure, and adhesion.
\newblock {\em Cell Motil Cytoskel}, 60(1):24--34, 2005.

\bibitem{Harris1980}
A.K. Harris, P.~Wild, and D.~Stopak.
\newblock Silicone rubber substrata: a new wrinkle in the study of cell
  locomotion.
\newblock {\em Science}, 208(4440):177, 1980.

\bibitem{Dembo1999}
M.~Dembo and Y.L. Wang.
\newblock Stresses at the cell-to-substrate interface during locomotion of
  fibroblasts.
\newblock {\em Biophysical journal}, 76(4):2307--2316, 1999.

\bibitem{Butler2002}
J.P. Butler, I.M. Toli{\'c}-N{\o}rrelykke, B.~Fabry, and J.J. Fredberg.
\newblock Traction fields, moments, and strain energy that cells exert on their
  surroundings.
\newblock {\em American Journal of Physiology-Cell Physiology},
  282(3):C595--C605, 2002.

\bibitem{Tan2003}
J.~L. Tan, J.~Tien, D.~M. Pirone, D.~S. Gray, K.~Bhadriraju, and C.~S. Chen.
\newblock Cells lying on a bed of microneedles: An apporach to isolate
  mechanical force.
\newblock {\em PMNAS}, 100:1484--1489, 2003.

\bibitem{Thery2010}
M.~Th{\'e}ry.
\newblock Micropatterning as a tool to decipher cell morphogenesis and
  functions.
\newblock {\em Journal of Cell Science}, 123(24):4201--4213, 2010.

\bibitem{Rape2011}
A.D. Rape, W.~Guo, and Y.~Wang.
\newblock The regulation of traction force in relation to cell shape and focal
  adhesions.
\newblock {\em Biomaterials}, 32(8):2043--2051, 2011.

\bibitem{deshpande2006}
V.S. Deshpande, R.M. McMeeking, and A.G. Evans.
\newblock A bio-chemo-mechanical model for cell contractility.
\newblock {\em Proceedings of the National Academy of Sciences},
  103(38):14015--14020, 2006.

\bibitem{Edwards2011}
C.M. Edwards and U.S. Schwarz.
\newblock Force localization in contracting cell layers.
\newblock {\em Physical Review Letters}, 107(12):128101, 2011.

\bibitem{Banerjee2011}
S.~Banerjee and M.C. Marchetti.
\newblock Substrate rigidity deforms and polarizes active gels.
\newblock {\em EPL (Europhysics Letters)}, 96:28003, 2011.

\bibitem{lemmon2010}
C.A. Lemmon and L.H. Romer.
\newblock A predictive model of cell traction forces based on cell geometry.
\newblock {\em Biophysical journal}, 99(9):L78, 2010.

\bibitem{Torres2012}
P.G. Torres, IB~Bischofs, and US~Schwarz.
\newblock Contractile network models for adherent cells.
\newblock {\em Physical Review E}, 85(1):011913, 2012.

\bibitem{Lemmon2005}
C.A. Lemmon, N.J. Sniadecki, S.A. Ruiz, J.L. Tan, L.H. Romer, and C.S. Chen.
\newblock Shear force at the cell-matrix interface: enhanced analysis for
  microfabricated post array detectors.
\newblock {\em Mechanics \& chemistry of biosystems: MCB}, 2(1):1, 2005.

\bibitem{Barziv1999}
R.~Bar-Ziv, T.~Tlusty, E.~Moses, S.A. Safran, and A.~Bershadsky.
\newblock Pearling in cells: a clue to understanding cell shape.
\newblock {\em Proc Natl Acad Sci USA}, 96(18):10140--10145, 1999.

\bibitem{Bischofs2009}
I.B. Bischofs, S.S. Schmidt, and U.S. Schwarz.
\newblock Effect of adhesion geometry and rigidity on cellular force
  distributions.
\newblock {\em Phys Rev Lett}, 103(4):48101, 2009.

\bibitem{Banerjee2012b}
S.~Banerjee and L.~Giomi.
\newblock Polymorphism and bistability in adherent cells.
\newblock {\em arXiv preprint arXiv:1209.4004}, 2012.

\bibitem{Liverpool2009}
TB~Liverpool, M.C. Marchetti, J.F. Joanny, and J.~Prost.
\newblock Mechanical response of active gels.
\newblock {\em EPL (Europhysics Letters)}, 85(1):18007, 2009.

\bibitem{Banerjee2011b}
S.~Banerjee, T.B. Liverpool, and M.C. Marchetti.
\newblock Generic phases of cross-linked active gels: Relaxation, oscillation
  and contractility.
\newblock {\em EPL (Europhysics Letters)}, 96(5):58004, 2011.

\bibitem{Legant2013}
W.R. Legant, C.K. Choi, J.S. Miller, L.~Shao, L.~Gao, E.~Betzig, and C.S. Chen.
\newblock Multidimensional traction force microscopy reveals out-of-plane
  rotational moments about focal adhesions.
\newblock {\em Proceedings of the National Academy of Sciences},
  110(3):881--886, 2013.

\bibitem{Banerjee2012}
S.~Banerjee and M.C. Marchetti.
\newblock Contractile stresses in cohesive cell layers on finite-thickness
  substrates.
\newblock {\em Phys Rev Lett}, 109:108101, Sep 2012.

\bibitem{Chen1997}
C.S. Chen, M.~Mrksich, S.~Huang, G.M. Whitesides, and D.E. Ingber.
\newblock Geometric control of cell life and death.
\newblock {\em Science}, 276(5317):1425--1428, 1997.

\bibitem{Thery2006}
M.~Th{\'e}ry, A.~P{\'e}pin, E.~Dressaire, Y.~Chen, and M.~Bornens.
\newblock Cell distribution of stress fibres in response to the geometry of the
  adhesive environment.
\newblock {\em Cell Motil Cytoskel}, 63(6):341--355, 2006.

\bibitem{Mertz2012}
A.F. Mertz, S.~Banerjee, Y.~Che, G.K. German, Y.~Xu, C.~Hyland, M.C. Marchetti,
  V.~Horsley, and E.R. Dufresne.
\newblock Scaling of traction forces with the size of cohesive cell colonies.
\newblock {\em Phys Rev Lett}, 108(19):198101, 2012.

\bibitem{Chopra2011}
A.~Chopra, E.~Tabdanov, H.~Patel, P.A. Janmey, and J.Y. Kresh.
\newblock Cardiac myocyte remodeling mediated by n-cadherin-dependent
  mechanosensing.
\newblock {\em Am J Physiol-Heart C}, 300(4):H1252--H1266, 2011.

\bibitem{Ghibaudo2008}
M.~Ghibaudo, A.~Saez, L.~Trichet, A.~Xayaphoummine, J.~Browaeys, P.~Silberzan,
  A.~Buguin, and B.~Ladoux.
\newblock Traction forces and rigidity sensing regulate cell functions.
\newblock {\em Soft Matter}, 4(9):1836--1843, 2008.

\bibitem{Wakatsuki2003}
T.~Wakatsuki, R.B. Wysolmerski, and E.L. Elson.
\newblock Mechanics of cell spreading: role of myosin ii.
\newblock {\em J Cell Sci}, 116(8):1617--1625, 2003.

\bibitem{Wang2002}
N.~Wang, I.M. Toli{\'c}-N{\o}rrelykke, J.~Chen, S.M. Mijailovich, J.P. Butler,
  J.J. Fredberg, and D.~Stamenovi{\'c}.
\newblock Cell prestress. i. stiffness and prestress are closely associated in
  adherent contractile cells.
\newblock {\em American Journal of Physiology-Cell Physiology},
  282(3):C606--C616, 2002.

\bibitem{Tambe2011}
D.T. Tambe, C.C. Hardin, T.E. Angelini, K.~Rajendran, C.Y. Park,
  X.~Serra-Picamal, E.H. Zhou, M.H. Zaman, J.P. Butler, D.A. Weitz, et~al.
\newblock Collective cell guidance by cooperative intercellular forces.
\newblock {\em Nature materials}, 10(6):469--475, 2011.

\bibitem{roca2008}
P.~Roca-Cusachs, J.~Alcaraz, R.~Sunyer, J.~Samitier, R.~Farr{\'e}, and
  D.~Navajas.
\newblock Micropatterning of single endothelial cell shape reveals a tight
  coupling between nuclear volume in g1 and proliferation.
\newblock {\em Biophysical journal}, 94(12):4984--4995, 2008.

\bibitem{Heidemann2004}
S.R. Heidemann and D.~Wirtz.
\newblock Towards a regional approach to cell mechanics.
\newblock {\em Trends in cell biology}, 14(4):160--166, 2004.

\end{thebibliography}

\end{document}